\documentclass[nofootinbib,amsmath,amsfonts,twocolumn]{revtex4}
\usepackage{graphicx,xcolor}
\usepackage{longtable}
\usepackage{subfigure}

\newcommand{\eqr}[1]{Eq.~\eqref{#1}}

\newcommand{\secr}[1]{Sec.~\ref{#1}}

\newcommand{\figr}[1]{Fig.~\ref{#1}}

\newcommand{\vR}{{\mathbf{r}}}

\newcommand{\ic}{\scriptscriptstyle\mathrm{CIT}}
\newcommand{\ie}{\scriptscriptstyle\mathrm{MCV}}

\newcommand{\Hml}{\mathcal{W}}
\newcommand{\wave}{\mathrm{w}}

\begin{document}

\title{Local equilibrium and the second law of thermodynamics \\for irreversible systems with thermodynamic inertia}
\author{K.~S.~Glavatskiy}
\affiliation{School of Chemical Engineering, the University of Queensland, St Lucia QLD 4072, Australia}

\begin{abstract}
Validity of local equilibrium has been questioned for non-equilibrium systems which are characterized by delayed response. In particular, for systems with non-zero thermodynamic inertia, the assumption of local equilibrium leads to negative values of the entropy production, which is in contradiction with the second law of thermodynamics. In this paper we address this question by suggesting a variational formulation of irreversible evolution of a system with non-zero thermodynamic inertia. We introduce the Lagrangian, which depends on the properties of the normal and the so-called "mirror-image" systems. We show that the standard evolution equations, in particular the Maxwell-Cattaneo-Vernotte equation, can be derived from the variational procedure without going beyond the assumption of local equilibrium. We also argue, that the second law of thermodynamics should be understood as a consequence of the variational procedure and the property of local equilibrium. For systems with instantaneous response this leads to the standard requirement of the local instantaneous entropy production being always positive. However, if a system is characterized by delayed response, the formulation of the second law of thermodynamics should be altered. In particular, the quantity, which is always positive, is not the instantaneous entropy production, but the entropy production averaged over the period of the heat wave.
\end{abstract}

\maketitle



\section{Introduction}

In classical irreversible thermodynamics (CIT), heat propagation is described by Fourier's law, which states that the heat flux is proportional to the temperature gradient, $J =  -\lambda\nabla T$. This relation is in agreement with the second law of thermodynamics. Indeed, in case of heat conduction, the entropy production is proportional to the product of the heat flux and minus the gradient of the temperature \cite{deGrootMazur}. Thus, the entropy production is always positive (zero in equilibrium), which is one of the statements of the second law of thermodynamics.

However, Fourier's law of heat conduction is not the only known constitutive relation \cite{Van2012, Tamma1998}. In particular, some systems can exhibit delayed response \cite{Tzou} or even wave behavior \cite{Joseph1989}. A common extension of Fourier's law of heat conduction is the so-called Maxwell-Cattaneo-Vernotte (MCV) equation \cite{Maxwell1867, Cattaneo1948, Vernotte1958}
\begin{equation}\label{eq/01}
\tau\dot{J} + J = -\lambda\nabla T 
\end{equation}
where $\tau$ is a parameter with the dimensionality of time, and dot above a variable indicates the partial time derivative. The modification with respect to the standard Fourier equation is the term $\tau\dot{J}$. It can be viewed as the first term in the Taylor expansion of the flux $J(t+\tau)$ around $J(t)$, so the entire left hand side of \eqr{eq/01} represents the delayed response of the system at the time $t+\tau$ to the perturbation at the time $t$. Delayed response implies that the system has some sort of thermodynamic inertia, and the parameter $\tau$ is a mere of this inertia.

Existence of an additional term proportional to the temporal derivative in the evolution equation changes the nature of the temperature evolution. The system is no longer purely relaxing. In particular, if $\tau$ is large enough, the second term on the left hand side of \eqr{eq/01} may be neglected and we will obtain a wave equation. For not very large but non-zero values of $\tau$ the system behaves similarly to a damped harmonic oscillator \cite{Jou}. In particular, it relaxes towards equilibrium, and, depending on the values of the parameters $\tau$ and $\lambda$, may also exhibit oscillating behavior. Another important consequence of existence of the first term in \eqr{eq/01} is the restriction on the speed of heat propagation. In a system described by Fourier's law, a thermal perturbation propagates with infinite speed, which is one of the major critics of CIT. However, in the system described by \eqr{eq/01} this speed is finite and is proportional to $1/\sqrt{\tau}$. The speed of heat propagation diverges in the limit of vanishing $\tau$ and becomes infinite for the system described by Fourier's law.

It is a common understanding, that \eqr{eq/01}, being treated in the context of CIT, leads to violation of the second law of thermodynamics \cite{Jou}. It can be shown that the classical specific entropy oscillates with the course of time, if the system is described by \eqr{eq/01}. In particular, classical specific entropy can decrease with time, which contradicts to the requirement of the entropy production being always positive. To overcome this issue, there have been developed a theory known under the name "extended irreversible thermodynamics" (EIT) \cite{Lebon2015}. The key difference of EIT with respect to CIT is the relation to the \textit{local equilibrium} hypothesis. It is assumed in CIT that the thermodynamic relations formulated for equilibrium macroscopic systems are also valid in non-equilibrium, if applied locally in space and time. In other words, every small sub-volume of a system for a given moment of time is assumed to reach it's equilibrium, such that there exist \textit{local equilibrium}. Mathematically this is formulated such that the specific entropy is a function of the local values of the state variables (e.g. the local densities) only. In contrast, EIT assumes that in non-equilibrium the specific entropy may in addition depend on the fluxes as independent variables. Therefore, the local thermodynamic relations in non-equilibrium are not the same as in equilibrium, and the system is said to be not in local equilibrium. The important consequence of considering the specific entropy to be dependent on both, the state variables and the fluxes, is that the entropy production is always positive. In addition, dependence of the fluxes leads to the equation of the MCV type \eqref{eq/01}. Thus, consistency with the second law of thermodynamics is restored. 

In the earlier work \cite{Glavatskiy2015} we have shown that the second law of thermodynamics can be viewed not as an independent law, but as a \textit{consequence} of a variational procedure, which minimizes a certain thermodynamic "action". The important part of that procedure was the assumption of local equilibrium. It was possible to derive the classical force-flux relations, one of which had the form of Fourier's law of heat conduction. In view of that, it is interesting to realize, whether a similar procedure can result in a force-flux relation of the MCV type \eqref{eq/01}. Furthermore, it is interesting to understand whether local equilibrium is sufficient to describe the systems, which are governed by \eqr{eq/01}. In this paper we will address these questions. We will show that the property of local equilibrium is, in fact, \textit{sufficient} for the system to exhibit a MCV-like behavior. We will also address the issue of contradiction between the local equilibrium hypothesis and the second law of thermodynamics. In particular, we will show, that the formulation of the second law of thermodynamics, being considered as a consequence of the variational procedure, should be modified. 

Variational methods in irreversible thermodynamics have mostly been considered for restricted sets of conditions \cite{Schechter, Glavatskiy2015}. In particular, they have mostly been formulated either for the constant values of the transport coefficients or for stationary conditions. The celebrated principle of minimum total entropy production [p. 45]\cite{deGrootMazur}, formulated by Prigogine, states that in stationary states the evolution of the system with constant transport coefficients is such that the total entropy production is minimum. There have been other variational formulations of irreversible thermodynamics with various restrictions \cite{Glavatskiy2015, VanVariation}. Not much work has been done for the systems, which possess thermodynamic inertia. The systems of heat wave propagation, which are purely reversible, allow a standard variational formulation \cite{MorseFeshbach}. The MCV-like systems can be viewed as waves with dissipation, and the dissipation can be included by introducing an exponentially decaying factor in the Lagrangian \cite{Sieniutycz1984}. Such variational description can not be considered general enough, since the Lagrangian explicitly contains a particular solution. In addition, it is not symmetric with respect to time reversal, and therefore is not compatible with microscopic time reversibility. 

In our earlier work \cite{Glavatskiy2015} we formulated a variational procedure, which allows one to \textit{derive} the force-flux relations and the second law of thermodynamics for the systems without thermodynamic inertia. We suggested the Lagrangian which is symmetric with respect to time reversal and does not contain explicitly nether the solutions, nor the evolution equations. The variational principle is formulated for the so-called extended system, which consists of the normal system, the so-called mirror-image system, and the equilibrium system. Evolution of the normal system is characterized by the normal second law of thermodynamics, which states that during the evolution the entropy production is always positive, $\sigma >0$. In contrast, evolution of the mirror-image system is characterized by the negative entropy production, $\sigma^{*} <0$, which can be viewed as the mirror-image second law of thermodynamics. Finally, the equilibrium system is required as a reference system for the other two. The description of the irreversible evolution of such system is completely equivalent to the one, given by the standard CIT \cite{deGrootMazur}. In this paper we extend this analysis to the systems with non-zero thermodynamic inertia. 

The paper is organized as follows. In \secr{sec/CIT} we briefly mention the important steps of the variational procedure for the system governed by CIT. In this way we also introduce the notation and the important principles. In \secr{sec/MCV} we extend the variational procedure to the systems with thermodynamic inertia. We present the Lagrangian, which leads to MCV-like equations of evolution. We also present a conserved quantity, which is an analogue of the Hamiltonian in classical mechanics. Conservation of this Hamiltonian during evolution is important for understanding the entropy transformations in the system. In \secr{sec/Resistless} we consider the specific case of the MCV-like system, the thermal wave propagation. Such system has no dissipation but still has thermodynamic inertia. It is therefore convenient to view such system as a reference system for our analysis. In \secr{sec/Complex} we introduce the coordinates, in which evolution of the system has simplest form and is easy to analyze. We discuss the entropy transformation in the system with both non-zero thermodynamic inertia and dissipation in \secr{sec/Entropy}. In particular, we give a new formulation of the second law of thermodynamics, which does not contradict to local equilibrium. Finally, in \secr{sec/Discussion} we discuss the similarities between the presented theory and the known and developed fields of physics, in particular, classical mechanics and electrical engineering.

\section{Classical irreversible thermodynamics}\label{sec/CIT}

Irreversible evolution of the system, which is described by the Fourier-like law between the thermodynamic force and the thermodynamic flux is derived from the principle of stationary action, with the Lagrangian given by
\begin{equation}\label{eq/02}
\widetilde{L}_{\ic} = -\ell(\varphi^{eq})\,\nabla\varphi\cdot\nabla\varphi^{*} - \frac{1}{2}\left(\varphi\dot{\rho}^{*} - \varphi^{*}\dot{\rho}\right) 
\end{equation}
Here and further the quantities with the tilde $\widetilde{\phantom{o}}$ will denote the properties related to the extended system, i.e. to the all three systems together, the normal one, the mirror-image one and the equilibrium one. The Lagrangian of the extended system depends on the potential $\varphi$ and the material density $\rho$, which are related by an equation of state $\rho = \rho_{_{EOS}}(\varphi)$. In the case of heat conduction, the potential is $1/T$, where $T$ is the temperature, while the density is the internal energy density. In the case of diffusion the potential is $-\mu/T$, where $\mu$ is the chemical potential, while the density is the density of the component. In the case of electric conduction the potential is $-\phi/T$, where $\phi$ is the electric potential, while the density is the charge density. The important property of the Lagrangian \eqref{eq/02} is that it is symmetric in time. The actual evolution of the system, which makes the irreversible action $\int L\,d\vR\,dt$ to reach its extremum, is described by the solution of the Euler-Lagrange equations. 

The Euler-Lagrange equations represent the pair of the force-flux relations, in the normal system and in the mirror-image system:
\begin{equation}\label{eq/03}
\begin{array}{rl}
J =& \ell\,\nabla\varphi
\\\\
J^{*} =& -\ell\,\nabla\varphi^{*}
\end{array}
\end{equation}
The coefficient $\ell$ is the ordinary phenomenological coefficient \cite{deGrootMazur}. It is positive. It is an even function with respect to time reversal. And it is independent of either $\varphi$ or $\varphi^{*}$. According to the Green-Kubo relations, it is equal to the equilibrium time-correlation function of corresponding microscopic fluxes. For not far from equilibrium perturbations we also have 
\begin{equation}\label{eq/05}
\ell(\varphi) \approx \ell(\varphi^{eq}) \approx \ell(\varphi^{*})
\end{equation}

The actual evolution of the system obeys the property of local equilibrium. This, in particular, means that non-equilibrium equation of state has the same form as in equilibrium, $\rho(\vR,t) = \rho_{_{EOS}}(\varphi(\vR,t))$ and $\rho^{*}(\vR,t) = \rho_{_{EOS}}(\varphi^{*}(\vR,t))$, where the function $\rho_{_{EOS}}$ is defined by the equilibrium relation $\rho^{eq} = \rho_{_{EOS}}(\varphi^{eq})$. Because of this, we can write that 
\begin{equation}\label{eq/04}
\frac{\partial \rho^{*}(\vR,t)}{\partial \varphi^{*}(\vR,t)} = 
\frac{\partial \rho^{eq}}{\partial \varphi^{eq}}=
\frac{\partial \rho(\vR,t)}{\partial \varphi(\vR,t)} \equiv -\chi
\end{equation}
where for further convenience we introduced a positive quantity $\chi$, which we will call a capacity. Indeed, in the case of heat conduction the potential $\varphi \equiv 1/T$ and the material density $\rho \equiv u$ is the internal energy density. The derivative $\partial u /\partial (1/T) = -T^2\,C$, where $C$ is the heat capacity. The heat capacity is positive, which makes $\chi$ to be positive and proportional to the ordinary heat capacity. A similar identifications can be made in the case of diffusion and electric conduction. 

We should emphasize, that we require the property of local equilibrium to hold only for the actual evolution trajectory, i.e. the  one, which satisfies the Euler-Lagrange equations. This means that \eqr{eq/04} and \eqr{eq/05} are correct only for the solution of the Euler-Lagrange equations, but not for any other evolution trajectory. In other words, the assignments \eqref{eq/04} and \eqref{eq/05} should not be made directly in the Lagrangian, but only in the resulting Euler-Lagrange equations, i.e. after the variation of the Lagrangian is performed. Because of this $\ell$ and $\chi$ may have any dependence on the equilibrium state functions (e.g. the temperature), and do not have to be constants.

We have also shown that the above procedure allows one to \textit{derive} the second law of thermodynamics. The variational principle implies that in the course of evolution there exists a quantity $\widetilde{\Hml}$, which is the analogue of the Hamiltonian density in classical mechanics: its integral over the volume of the system is conserved with the time 
\begin{equation}\label{eq/06}
\frac{d}{dt}\int_{V}{\widetilde{\Hml}\,dV} = 0
\end{equation}
The Hamiltonian density for the Lagrangian \eqref{eq/02} has the form 
\begin{equation}\label{eq/07}
\widetilde{\Hml}_{\ic} = \ell(\varphi^{eq})\,\nabla\varphi\cdot\nabla\varphi^{*}
\end{equation}
which implies that the entropy productions in the normal and the mirror-image systems have opposite signs. Indeed, their product is negative,  $\sigma\sigma^{*} = -\widetilde{\Hml}_{\ic}^2$, and the entropy production of the normal system is always positive, $\sigma > 0$.

\section{Maxwell-Cattaneo-Vernotte equation}\label{sec/MCV}

Let us consider the following Lagrangian
\begin{equation}\label{eq/11}
\widetilde{L}_{\ie} = m(\varphi^{eq})\,\dot{\rho}\,\dot{\rho}^{*} - \ell(\varphi^{eq})\,\nabla\varphi\cdot\nabla\varphi^{*}
-\frac{1}{2}\left(\varphi\dot{\rho}^{*} - \varphi^{*}\dot{\rho}\right)  
\end{equation}
It differs from $\widetilde{L}_{\ic}$ by an additional term, which is proportional to the product of the time rates of change of the material densities of in the normal and the mirror-image systems. As we will see further, the proportionality coefficient $m$ is a mere of the thermodynamic inertia of the system. Just like the coefficient $\ell$, $m$ is positive and is an even function with respect to time reversal, and also is independent of either $\varphi$ or $\varphi^{*}$. The Lagrangian \eqref{eq/11} is symmetric in time, just like the Lagrangian \eqref{eq/02}. 

Just like for the classical irreversible thermodynamics, we assume that the actual evolution of the system obeys the property of local equilibrium. This means that \eqr{eq/04} and \eqr{eq/05} are still true for the trajectory, which makes the irreversible action $\int L_{\ie}\,d\vR\,dt$ to be extremal. In addition, we require for the extremal trajectory not far from equilibrium that the inertial function $m$ satisfies
\begin{equation}\label{eq/12}
m(\varphi) \approx m(\varphi^{eq}) \approx m(\varphi^{*})
\end{equation}

The Lagrangian \eqref{eq/11} depends on the first-order spatial and temporal derivatives of the state functions. The Euler-Lagrange equations have therefore the following form
\begin{equation}\label{eq/13}
\begin{array}{rl}
\displaystyle\frac{\partial \widetilde{L}_{\ie}}{\partial\varphi} -
\frac{\partial}{\partial t}\frac{\partial \widetilde{L}_{\ie}}{\partial \dot{\varphi}} -
\nabla\cdot\frac{\partial \widetilde{L}_{\ie}}{\partial \nabla\varphi} &= 0 
\\\\
\displaystyle\frac{\partial \widetilde{L}_{\ie}}{\partial\varphi^{*}} -
\frac{\partial}{\partial t}\frac{\partial \widetilde{L}_{\ie}}{\partial \dot{\varphi}^{*}} -
\nabla\cdot\frac{\partial \widetilde{L}_{\ie}}{\partial \nabla\varphi^{*}} &= 0 
\end{array}
\end{equation}
Evaluating the partial derivatives and taking into account \eqr{eq/04}, \eqr{eq/05} and \eqr{eq/12} we obtain the evolution equations in the normal and the mirror-image systems
\begin{equation}\label{eq/14}
\begin{array}{rl}
\tau\ddot{\rho} + \dot{\rho} + \nabla\cdot(\ell\,\nabla\varphi) &=0
\\\\
\tau\ddot{\rho}^{*} - \dot{\rho}^{*} + \nabla\cdot(\ell\,\nabla\varphi^{*}) &=0
\end{array}
\end{equation}
where we introduced 
\begin{equation}\label{eq/14a}
\tau(\varphi^{eq}) \equiv \chi\,m
\end{equation}
The parameter $\tau$ is positive and has dimensionality of time. The terms $\tau\ddot{\rho}$ and $\tau\ddot{\rho}^{*}$ represent delayed responses of the material densities to the perturbations $\nabla\varphi$ and $\nabla\varphi^{*}$. Thus, the parameter $\tau$ may be identified with the delay time of the system's response, which is a consequence of non-zero thermodynamic inertia. The delay time is proportional to the inertial parameter $m$ and characterizes thermodynamic inertia of the system. In particular, when the inertial parameter $m=0$, the system has no inertia, and the evolution equations become equivalent to the force-flux relations of the classical irreversible thermodynamics.

We next combine \eqr{eq/14} with the balance equations for the normal and the mirror image convection-free systems. There the sum of the divergence of the material flux and the time rate of change of the material density is equal to zero for both, the normal system, $\nabla J + \dot{\rho} = 0$, and the mirror-image system, $\nabla J^{*} + \dot{\rho}^{*} = 0$. It follows therefore that
\begin{equation}\label{eq/15}
\begin{array}{rl}
\tau\dot{J} + J - \ell\,\nabla\varphi &= 0
\\\\
\tau\dot{J}^{*} - J^{*} - \ell\,\nabla\varphi^{*} &= 0
\end{array}
\end{equation}
\eqr{eq/15} are the force-flux relations for an irreversible system with non-zero thermodynamic inertia. For heat conduction the potential $\varphi \equiv 1/T$ and it is easy to see that the first of \eqr{eq/15} is the MCV equation \eqref{eq/01} with the thermal conductivity related to the phenomenological coefficient in a standard way, $\lambda = \ell/T^{2}$.

It follows therefore that evolution of the inertial system, which is described by the MCV-like equation, can be derived from the variational principle. The important step in the derivation is the property of local equilibrium, which allows one to apply \eqr{eq/04}, \eqr{eq/05} and \eqr{eq/12} for the extremal evolution trajectory. The variational procedure for the inertial system is the same as the one for the inertia-less system, introduced in \cite{Glavatskiy2015} and outlined in the previous section. The difference between the evolution of the inertial and inertial-less system is due to the additional term $m\dot{\rho}\dot{\rho}^{*}$ in the Lagrangian $\widetilde{L}_{\ie}$ compared to the Lagrangian $\widetilde{L}_{\ic}$. We shall call this term the kinetic entropy production. When the inertial parameter $m=0$, the Lagrangian $\widetilde{L}_{\ie}$ becomes equivalent to the Lagrangian $\widetilde{L}_{\ic}$, and the evolution of the system follows the classical Fourier-like equation of heat conduction. 

As in the case of inertia-less system, the variational procedure for the inertial system implies that in the course of evolution there exists a Hamiltonian density $\widetilde{\Hml}_{\ic}$, integral of which over the volume of the system is conserved with time, so that \eqr{eq/06} holds for $\widetilde{\Hml}_{\ic}$. The Hamiltonian density is $\widetilde{\Hml} \equiv -\widetilde{L} + \dot{\varphi}(\partial \widetilde{L}/\partial \dot{\varphi}) + \dot{\varphi}^{*}(\partial \widetilde{L}/\partial \dot{\varphi}^{*})$. In the case of Lagrangian \eqref{eq/11}, this results in the following expression
\begin{equation}\label{eq/16}
\widetilde{\Hml}_{\ie} = \ell(\varphi^{eq})\,\nabla\varphi\cdot\nabla\varphi^{*} + m(\varphi^{eq})\,\dot{\rho}\,\dot{\rho}^{*} 
\end{equation}
The Hamiltonian density for the inertial system differs from the Hamiltonian density for the inertia-less system by the kinetic entropy production. The conserved quantity in the inertial system, the volume integral over the Hamiltonian density, contains therefore a kinetic contribution.

\section{Resistless evolution}\label{sec/Resistless}

Depending on the values of the parameters $\tau$ and $\ell$, the system may exhibit specific behavior. In particular, when $\tau$ is very large, the second term in \eqr{eq/15} may be neglected, and we obtain two wave equations. Wave propagation is a conservative process, so the limit of large $\tau$ corresponds to a system without dissipation, i.e. resistless system. However, the parameter $\tau$ characterizes delayed response of the system to external perturbation, which seems to be unrelated to the dissipative properties of the system and it might be not convenient to use $\tau$ as a mere of dissipation in the system. We formalize this observation, we shall view the resistless system from a different perspective.

In particular, let us introduce the resistivity $r(\varphi^{eq}) \equiv 1/\ell$. Furthermore, let us introduce the heat wave factor $\wave(\varphi^{eq})$ and can the heat wave speed $c(\varphi^{eq})$:
\begin{equation}\label{eq/21}
\begin{array}{rl}
\wave^2 &\equiv \displaystyle \frac{\ell}{m} = \frac{1}{\chi}\,\frac{\ell}{\tau} 
\\\\
c^2 &\equiv \displaystyle \frac{\wave^2}{\chi^2} = \frac{1}{\chi^{3}}\,\frac{\ell}{\tau} 
\end{array}
\end{equation}
As wave propagation is a conservative process, its speed may not depend on the irreversible characteristics, in particular the phenomenological transfer coefficient $\ell$. We may expect therefore $m$ to be proportional to $\ell$. Furthermore, we may expect that it is $\wave$ (or $c$) that reflects the "true" inertial characteristics of the system, while $m$ is a combination of the irreversible characteristic $\ell$ and the inertial characteristic $\wave$ (or $c$). 

With the new quantities the Lagrangian \eqref{eq/11} can be written as 
\begin{equation}\label{eq/24}
\begin{array}{rl}
\displaystyle \widetilde{L}_{\ie} = \ell(\varphi^{eq})\,\left[\vphantom{\frac{1}{\wave^2}}\right.
\!&\!\displaystyle \frac{1}{\wave^2}\,\dot{\rho}\,\dot{\rho}^{*} - \nabla\varphi\cdot\nabla\varphi^{*}
\\\\
\!&\!\displaystyle \left.-\frac{1}{2}\,r(\varphi^{eq})\,\left(\varphi\dot{\rho}^{*} - \varphi^{*}\dot{\rho}\right)  \right]
\end{array}
\end{equation}
It leads to the same irreversible evolution equations \eqref{eq/14} and results in the same MCV-type equation \eqref{eq/15}. The wave equation is obtained when $r = 0$. Indeed, in this case the Lagrangian becomes 
\begin{equation}\label{eq/25}
\widetilde{L}_{\wave} = \ell(\varphi^{eq})\left[\frac{1}{\wave^2}\,\dot{\rho}\,\dot{\rho}^{*} - \nabla\varphi\cdot\nabla\varphi^{*}\right]
\end{equation}
If the coefficients are constant, the minimizing the Lagrangian $\widetilde{L}_{\wave}$ results in the following wave equations, which have the same form for both, the normal system and the mirror-image system:
\begin{equation}\label{eq/23}
\begin{array}{rl}
\ddot{\varphi} =& c^2\,\nabla^2\varphi
\\\\
\ddot{\varphi}^{*} =& c^2\,\nabla^2\varphi^{*}
\end{array}
\end{equation}
In other words, the resistless evolution corresponds to the wave propagation. We can see, that considering a system with a large relaxation time $\tau$ is equivalent to considering a system with a small resistivity $r$. In particular, when $r = 0$, $\tau$ is infinitely large. 

The Hamiltonian density $\widetilde{\Hml}_{\wave}$ for the resistless system has the same form as for an ordinary MCV-like system, \eqr{eq/16}:
\begin{equation}\label{eq/32}
\widetilde{\Hml}_{\wave} = m(\varphi^{eq})\,\dot{\rho}\,\dot{\rho}^{*} + \ell(\varphi^{eq})\,\nabla\varphi\cdot\nabla\varphi^{*}
\end{equation}
This means, in particular, that if a resistless system has the same initial and boundary conditions as the corresponding MCV-like system, then the conserved Hamiltonian has the same value in both systems. 

It is interesting to observe, that the same wave equations can be obtained from the Lagrangians, which involve the state variables from only one system, either the normal one, or the mirror-image one. Indeed, consider the following single-system Lagraninans 
\begin{equation}\label{eq/26}
\begin{array}{rl}
L_{\wave} =& \displaystyle\frac{\ell(\varphi^{eq})}{2}\left[\frac{1}{\wave^2}\,|\dot{\rho}|^2 - |\nabla\varphi|^2\right]
\\\\
L_{\wave}^{*} =& \displaystyle\frac{\ell(\varphi^{eq})}{2}\left[\frac{1}{\wave^2}\,|\dot{\rho}^{*}|^2 - |\nabla\varphi^{*}|^2\right]
\end{array}
\end{equation}
The Lagrangian $L_{\wave}$ describes evolution of the normal system only. Similarly, the Lagrangian $L_{\wave}^{*}$ describes evolution of the mirror-image system only. Performing the variational procedure for each of this Lagrangians independently results in the wave equation. In case of the constant coefficients they have the form of \eqr{eq/23}. As expected, they have the same form in both, the normal system and the mirror-image system. One can conclude, that evolution of the extended resistless system is equivalent to the combined evolution of the normal resistless system and the mirror-image resistless system. In particular, the Lagrangian $L_{\wave} + L_{\wave}^{*}$ describe the same resistless system as the Lagrangian $\widetilde{L}_{\wave}$.

The single-system Hamiltonian densities for the normal and the mirror image systems have the following form
\begin{equation}\label{eq/33}
\begin{array}{rl}
\Hml_{\wave} =& \displaystyle \frac{m(\varphi^{eq})}{2}\,|\dot{\rho}|^2 + \frac{\ell(\varphi^{eq})}{2}\,|\nabla\varphi|^2
\\\\
\Hml_{\wave}^{*} =& \displaystyle  \frac{m(\varphi^{eq})}{2}\,|\dot{\rho}^{*}|^2 + \frac{\ell(\varphi^{eq})}{2}\,|\nabla\varphi^{*}|^2
\end{array}
\end{equation}
As $\widetilde{\Hml}_{\wave}$, they satisfy the conservation equation \eqref{eq/06}. It follows, in particular, that $\Hml_{\wave} + \Hml_{\wave}^{*} = \widetilde{\Hml}_{\wave}$. 

We see that evolution of the extended system, which is resistless, can be considered as a joint evolution of the normal and the mirror-image system, which evolve independently. In other words, resistless system is decoupled into the normal one and the mirror-image one. It follows therefore that the resistivity $r$ is a mere of coupling between the normal and the mirror-image system. When $r=0$ these two systems are completely decoupled, and each of them represent a conservative system with wave propagation. Nonzero $r$ introduces coupling between these system, which makes each of them irreversible. In particular, the volume integrals of $\Hml_{\wave}$ and $\Hml_{\wave}^{*}$ are no longer conserved. However, the extended system with nonzero resistivity still has a conserved quantity, the volume integral of $\widetilde{\Hml}_{\wave}$.

\section{Complex coordinates}\label{sec/Complex}

To understand the structure of, in particular, entropy transformation in the dissipative system with thermodynamic inertia, it is convenient to perform the analysis in complex space. We shall do that in this section.

\subsection{The generalized force}

A thermodynamic force, which drives the system away from equilibrium, is usually identified as $X \equiv \nabla\varphi$. In particular, for heat conduction it is $\nabla(1/T)$, for diffusion at constant temperature it is $-\nabla(\mu/T)$, for electric conduction at constant temperature it is $-\nabla(\phi/T)$. Let us consider a complex force, which is a generalization of the standard thermodynamic force. In particular, we introduce 
\begin{equation}\label{eq/41}
\Xi \equiv \displaystyle\nabla\varphi +  i\,\mathbf{e}_{\varphi}\,\frac{1}{\wave}\,\dot{\rho}\\\\
\end{equation}
where $i \equiv \sqrt{-1}$ is the imaginary unit and $\mathbf{e}_{\varphi}$ is the unit vector in the direction of the gradient of $\varphi$. Together with $\Xi$ we can consider the complex-conjugate force $\overline{\Xi}$, the mirror-image conjugate force $\Xi^{*}$, and their combination $\overline{\Xi}\vphantom{\Xi}^{*}$:
\begin{equation}\label{eq/42}
\begin{array}{rl}
\overline{\Xi}  					&\equiv \displaystyle\nabla\varphi -  i\,\mathbf{e}_{\varphi}\,\frac{1}{\wave}\,\dot{\rho}
\\\\
\Xi^{*} 							&\equiv \displaystyle\nabla\varphi^{*} -  i\,\mathbf{e}_{\varphi^{*}}\,\frac{1}{\wave}\,\dot{\rho}^{*}
\\\\
\overline{\Xi}\vphantom{\Xi}^{*} 	&= \displaystyle\nabla\varphi^{*} +  i\,\mathbf{e}_{\varphi^{*}}\,\frac{1}{\wave}\,\dot{\rho}^{*}
\end{array}
\end{equation}
Using this notation, the Hamiltonian density for the MCV-like system can be written as
\begin{equation}\label{eq/43}
\widetilde{\Hml}_{\ie} = \frac{\ell}{2}\left[\,\Xi\cdot\Xi^{*} + \overline{\Xi}\cdot\overline{\Xi}\vphantom{\Xi}^{*}\,\right]
\end{equation}
Furthermore for the resistless system,
\begin{equation}\label{eq/44}
\widetilde{\Hml}_{\wave} = \frac{\ell}{2}\left[\,\Xi\cdot\Xi^{*} + \overline{\Xi}\cdot\overline{\Xi}\vphantom{\Xi}^{*}\,\right]
\end{equation}
and
\begin{equation}\label{eq/45}
\begin{array}{rl}
\Hml_{\wave} 	 =& \displaystyle\frac{\ell}{2}\,\Xi\cdot\overline{\Xi}
\\\\
\Hml_{\wave}^{*} =& \displaystyle\frac{\ell}{2}\,\Xi^{*}\cdot\overline{\Xi}\vphantom{\Xi}^{*}
\end{array}
\end{equation}
while for the inertia-less system
\begin{equation}\label{eq/46}
\widetilde{\Hml}_{\ic} = \ell\,\Xi\cdot\Xi^{*}
\end{equation}

Let us mention first the inertia-less system. Evolution of such system was studied in \cite{Glavatskiy2015} and outlined in the second section of this paper. The generalized forces in this case is real, i.e. $\Xi = \overline{\Xi}$ and ${\Xi}^{*} = \overline{\Xi}\vphantom{\Xi}^{*}$. The conserved quantity is $\Xi\cdot\Xi^{*}$.

\begin{figure}[ht]
\centering
\subfigure[] %
{\includegraphics[scale=0.4]{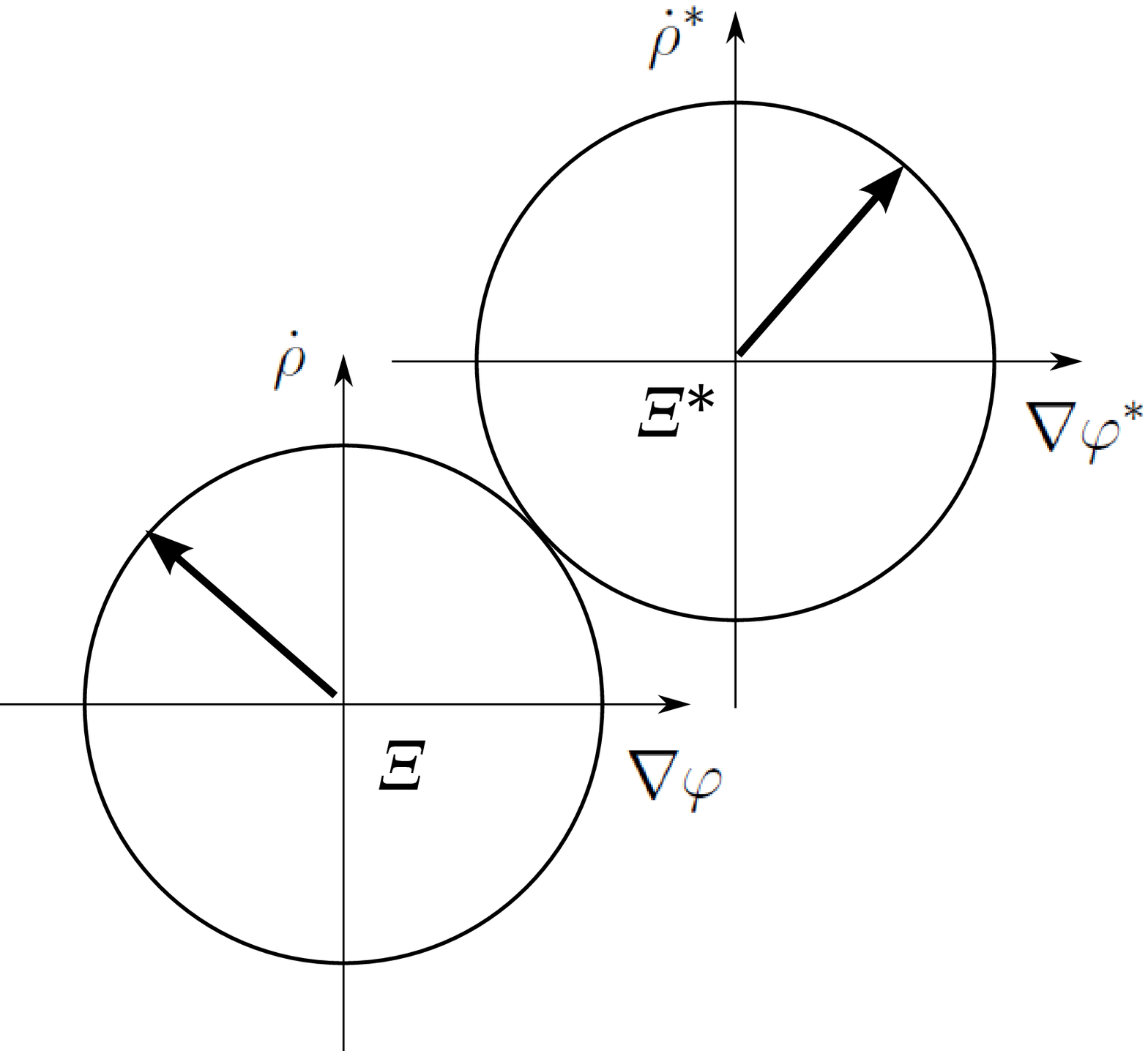}\label{fig-complex-resistless}}
~
\subfigure[] %
{\includegraphics[scale=0.4]{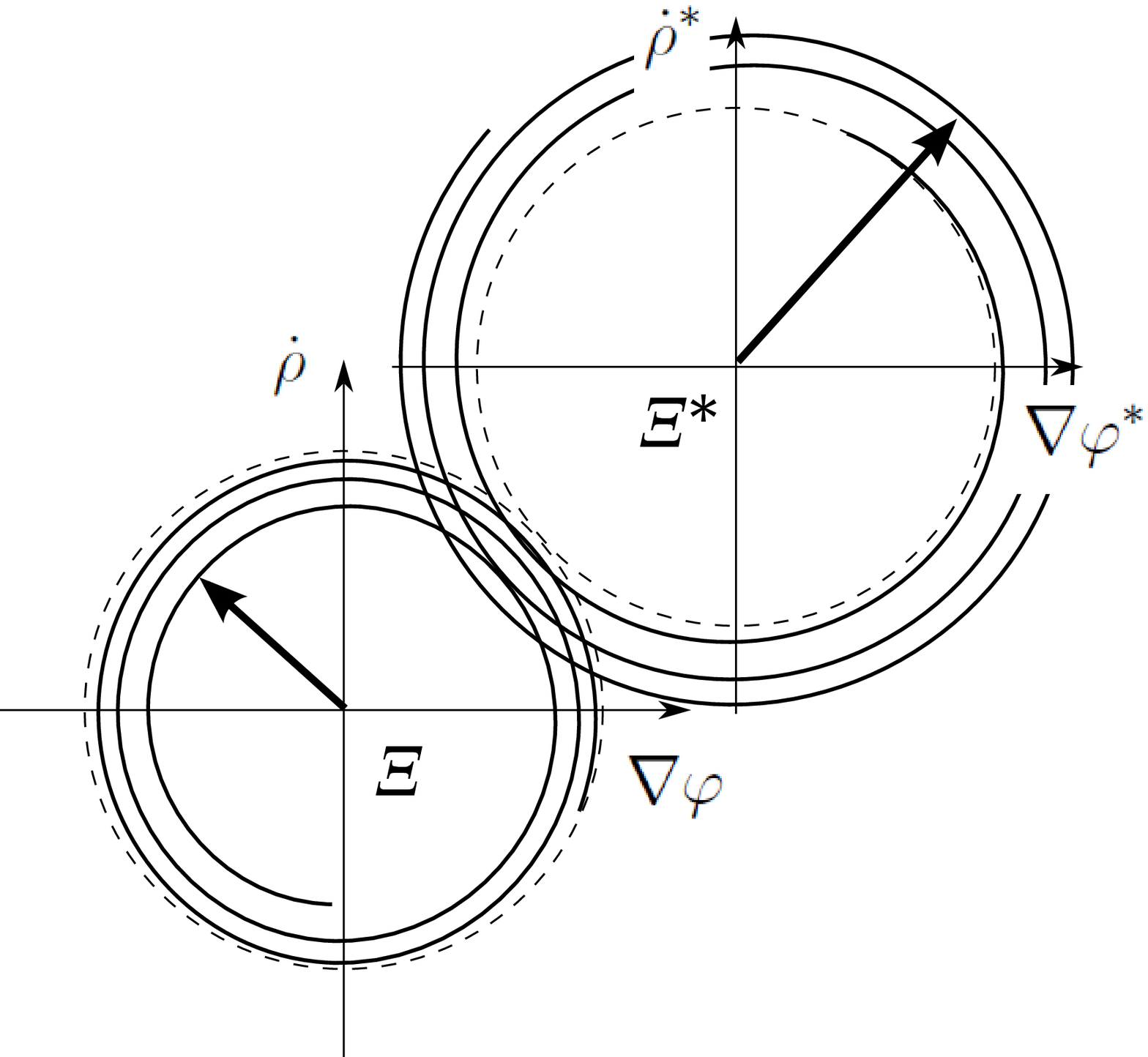}\label{fig-complex-dissipative}}
\caption{Evolution trajectory in the complex space. (a) Resistless system. (b) Dissipative system}\label{fig-complex}
\end{figure}

Let us consider now the resistless system. As we saw in the previous section, evolution of the normal and the mirror-image system is decoupled. Individual evolution of the normal system can be represented by a trajectory in the two-dimensional complex plane with the coordinates $|\nabla\varphi|$ and $\dot{\rho}$. Conservation of $\Hml_{\wave}$ means that the absolute value $|\Xi|^2$ of the generalized force is constant during evolution. Thus, the evolution trajectory of the normal system is a circle, and its evolution can be represented as rotation of the vector $\Xi$ in the complex plane. The same arguments are valid for the mirror-image system as well. We have therefore
\begin{equation}\label{eq/47}
\begin{array}{rl}
\displaystyle\frac{\dot{\rho}}{\wave|\nabla\varphi|} &= \tan (\omega t)
\\\\
\displaystyle\frac{\dot{\rho}^{*}}{\wave|\nabla\varphi^{*}|} &= -\tan (\omega t)
\end{array}
\end{equation}
where $\omega$ is the frequency and the sign of the right hand side is determined by the initial condition. On the other hand, evolution of the extended system, which contains the normal one and the mirror-image one, is represented by a trajectory in the four-dimensional complex space with the coordinates $|\nabla\varphi|$, $|\nabla\varphi^{*}|$ and $\dot{\rho}$, $\dot{\rho}^{*}$. Since the normal and the mirror-image systems are decoupled, evolution trajectory of the extended system is represented by two two-dimensional circles in the four-dimensional complex space, see \figr{fig-complex-resistless}. Each of this circle is a two-dimensional cross-section of the evolution trajectory of the extended system, which correspond to the pair of coordinates either $(|\nabla\varphi|; \dot{\rho})$ or $(|\nabla\varphi^{*}|; \dot{\rho}^{*})$. The radii of these circles are constant and equal to $|\Xi|$ and $|\Xi^{*}|$ respectively. Note, that for the resistless evolution $\widetilde{\Hml}_{\wave} = {\Hml}_{\wave} + {\Hml}^{*}_{\wave}$. This means that for the resistless evolution $\Xi^{*} = \overline{\Xi}$ and $\Xi = \overline{\Xi}\vphantom{\Xi}^{*}$. 

Finally, let us consider the extended system with non-zero resistivity and non-zero inertia. The normal and the mirror-image systems are coupled, which means that both $|\Xi|$ and $|\Xi^{*}|$ change. The two-dimensional cross-section of the evolution trajectory are now not circles, but spirals. The specific shape of the evolution trajectory depends on the relationships between the parameters of the system and the boundary conditions. In particular, in the case of a relaxation process, \eqr{eq/14} has a form of the equation for a damped oscillator with the damping ratio of the order $\chi\,c\,r$. When $\chi\,c\,r \lesssim 1$ the normal system experiences decay with oscillations, while the mirror-image system expands with oscillations, see \figr{fig-complex-dissipative}. When $\chi\,c\,r \gtrsim 1$ the corresponding relaxation and tightening processes happen without oscillations. In general, the vector of the generalized force changes such that the volume integral of the expression $\Xi\cdot\Xi^{*} + \overline{\Xi}\cdot\overline{\Xi}\vphantom{\Xi}^{*}$ remains constant. Also, $\Xi^{*} \neq \overline{\Xi}$ and $\Xi \neq \overline{\Xi}\vphantom{\Xi}^{*}$, as well as $\Xi \neq \overline{\Xi}$ and ${\Xi}^{*} \neq \overline{\Xi}\vphantom{\Xi}^{*}$. This means that in general $\widetilde{\Hml}_{\ie} \neq {\Hml}_{\wave} + {\Hml}^{*}_{\wave}$ and, in particular, the value, the volume integral of which remains constant, is neither $|\Xi|^2 + |\Xi^{*}|^2$, nor $\Xi\cdot\Xi^{*}$. 

\begin{figure}[ht]
\centering
\subfigure[] %
{\includegraphics[scale=0.5]{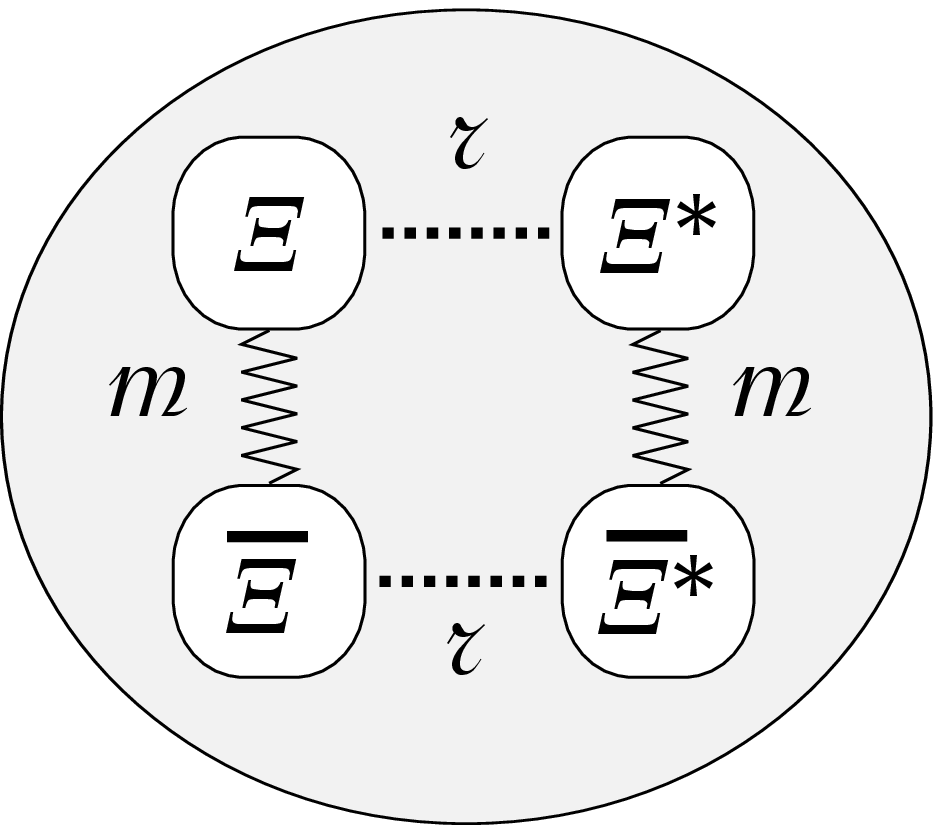}\label{fig-potentials-full}}
~~~~~
\subfigure[] %
{\includegraphics[scale=0.5]{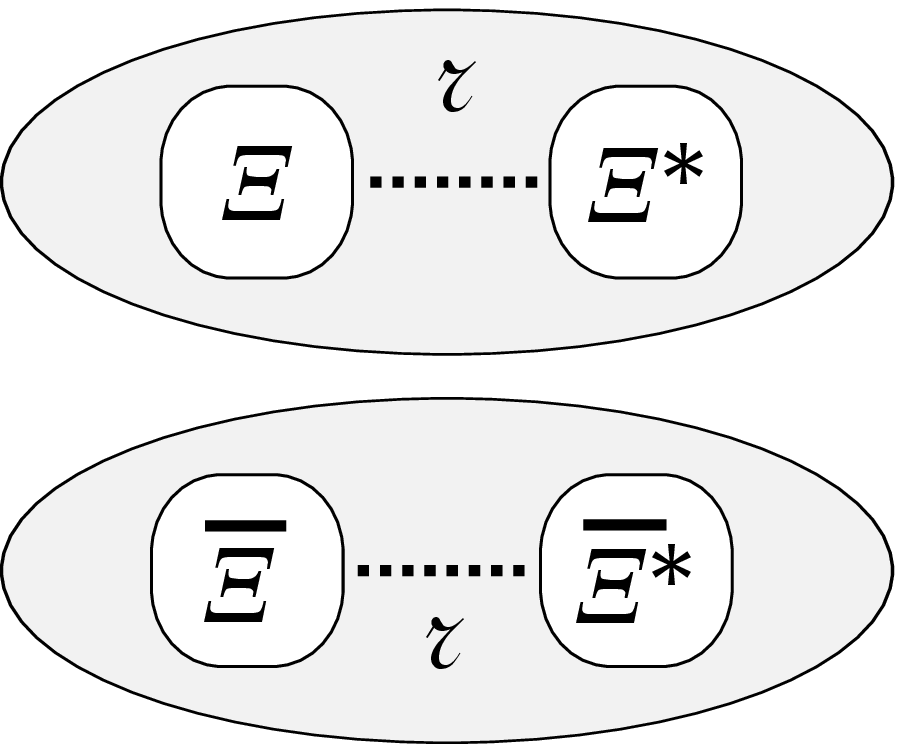}\label{fig-potentials-resistance}}%
~~~~~
\subfigure[] %
{\includegraphics[scale=0.5]{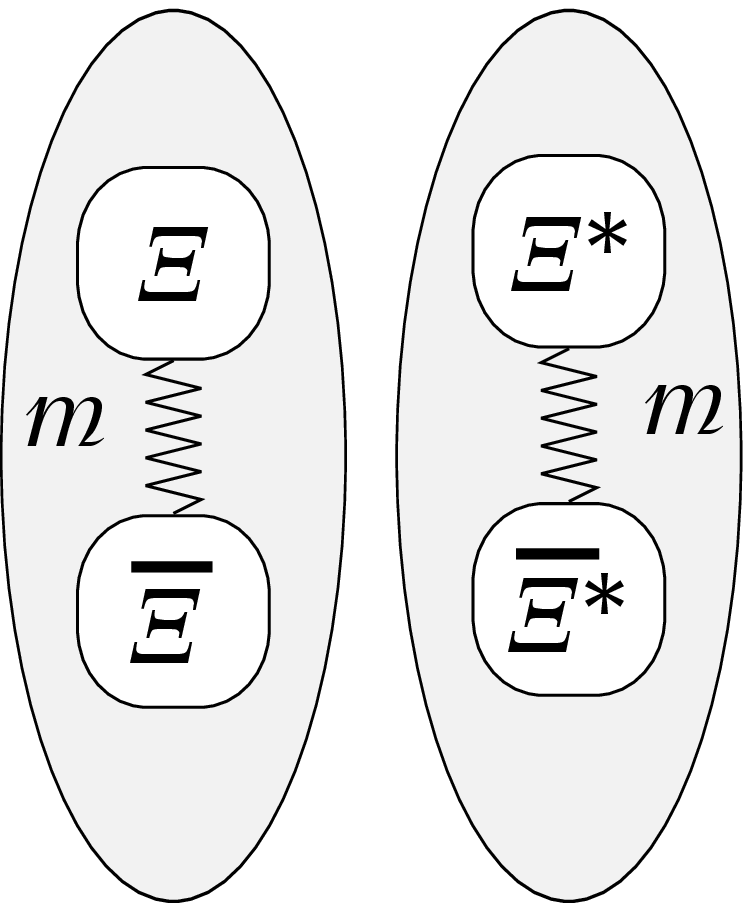}\label{fig-potentials-oscillation}}%
\caption{Interactions between the generalized forces in the (a) full system; (b) dissipative system; (c) resistless system. }\label{fig-potentials}
\end{figure}

The situations described above can also be illustrated by the diagrams in \figr{fig-potentials}. The resistless system (\figr{fig-potentials-oscillation}) allows independent wave propagation in both the normal and the mirror-image system. The inertia-less system (\figr{fig-potentials-resistance}) implies coupling between the normal system and the mirror-image one, as well as between their complex-conjugates. The full dissipative system with inertia (\figr{fig-potentials-full}) allows both coupling between the normal system and the mirror-image one, as well as wave propagation in either of them. The full system may not be decoupled in pairs like the resistless one or the inertia-less one.

\subsection{Impedance and force-flux relations}

Let us apply the Laplace transform to the MCV-like equation \eqref{eq/15}. The Laplace transform of a function $f$ is defined as $\widehat{f}(\kappa, \omega) \equiv \int_{0}^{\infty} \exp(-\varpi t)\, f\, dt$. Here $\varpi \equiv \kappa + i\omega$ is a complex frequency, where $\kappa$ is the relaxation parameter and $\omega$ is the wave frequency. The complex frequency characterizes the boundary (external) conditions, applied to the system (i.e. the temperature difference on the boundary, or the flux across the boundary). In particular, if the system is subjected to a steady-state boundary conditions which are periodic in time with the period $2\pi/\omega$, then $\varpi = i\omega$ is purely imaginary. In contrast, if the system is subjected to a steady-state fixed boundary conditions, then $\varpi = \kappa$ is purely real. A transient or a relaxation process is characterized by a complex frequency, with neither $\kappa$ nor $\omega$ are equal to zero. 
Laplace transform of \eqr{eq/15} has the following form:
\begin{equation}\label{eq/51}
\begin{array}{rl}
\widehat{J}(1+\tau\,\varpi) + J(0)&= \ell\,\nabla\widehat{\varphi} 
\\\\
\widehat{J}\vphantom{J}^{*}(-1+\tau\,\varpi) + J^{*}(0)&= \ell\,\nabla\widehat{\varphi}\vphantom{\varphi}^{*}
\end{array}
\end{equation}

The complex impedance is defined as a ratio of the Laplace transform of the force and the Laplace transform of the flux, when the initial values of the flux is taken to be equal to zero. We have therefore
\begin{equation}\label{eq/52}
\begin{array}{rl}
Z(\kappa, \omega) =& (1+\tau\,\kappa)\,r  + i \omega\,\tau\,r
\\\\
Z^{*} (\kappa, \omega) =& (-1+\tau\,\kappa)\,r + i \omega\,\tau\,r
\end{array}
\end{equation}
The complex impedance for each of the system consists of the real ans imaginary parts. The real part determines the active resistance and characterizes dissipation in the system. The relaxation parameter for the normal system is equal to $-r/\tau$, which indicates exponential decay of a perturbation. In contrast, the relaxation parameter of the mirror-image system is equal to $r/\tau$, which indicates an exponential growth of a perturbation. The imaginary part of the complex impedance determines the reactive resistance. It does not lead to dissipation or accumulation, but characterizes conservative transformations of the entropy in a heat wave. The reactive resistance is the same for both, the normal and the mirror-image systems.

Ignoring the initial condition, \eqr{eq/51} can be written as 
\begin{equation}\label{eq/53}
\begin{array}{rl}
\nabla\widehat{\varphi} =& Z(\kappa, \omega)\,\widehat{J}
\\\\
\nabla\widehat{\varphi}\vphantom{\varphi}^{*} =& Z^{*}(\kappa, \omega)\,\widehat{J}\vphantom{J}^{*}
\end{array}
\end{equation}
which is the force-flux relation for the irreversible system with thermodynamic inertia.

\section{Local equilibrium and the entropy production}\label{sec/Entropy}

The derivations in this paper rely on the property of local equilibrium. The property of local equilibrium is also crucial for classical irreversible thermodynamics. In particular, it is used when the Gibbs equation is formulated. This leads to the following expression for the local entropy production (which has the same form both for the normal and the mirror-image system):
\begin{equation}\label{eq/61}
\begin{array}{rl}
\sigma(\vR, t) =& J\cdot\nabla\varphi 
\\\\
\sigma^{*}(\vR,t) =& J^{*}\cdot\nabla\varphi^{*}
\end{array}
\end{equation}

If the system is inertia-less, the force-flux relations can be derived from the principle of stationary action \cite{Glavatskiy2015} and are given by \eqr{eq/03}. Substituting these relations in \eqr{eq/61}, leads to $\sigma(\vR, t) \geq 0$ and $\sigma^{*}(\vR, t) \geq 0$. The first of these relations is the statement of the second law of thermodynamics: the entropy production in the normal system is always positive, and reaches zero in equilibrium.

If the system has non-zero thermodynamic inertia $\tau$, the force-flux relations can still be derived from the principle of stationary action, which is the subject of this paper. These relations are given by \eqr{eq/15} instead of \eqr{eq/03}. Still, the derivation of \eqr{eq/15} implies the property of local equilibrium. Substituting \eqr{eq/15} in \eqr{eq/61} results in the entropy production, which cannot, in general, be represented as either a positive or a negative quantity. This leads to apparent violation of the second law of thermodynamics for the system with thermodynamic inertia.

There are two different conclusions which can be made on the basis of this observation: either the second law of thermodynamics should be reformulated, or the property of local equilibrium should be questioned. The latter conclusion is employed by extended irreversible thermodynamics \cite{Jou}. In particular, the non-equilibrium Gibbs relation is modified such that the non-equilibrium local entropy depends not only on the state variables, but also on the irreversible fluxes. This results in the expression for the entropy production, which is always positive in non-equilibrium.

The alternative conclusion, which requires revisiting the second law of thermodynamics, may sound unreasonable. Still, the analysis in this paper shows, that the MCV-like force-flux equations \eqref{eq/15} can be derived rigorously within the assumption of local equilibrium. The property of local equilibrium is \textit{sufficient} for  deriving the MCV-like force-flux equations. In other words, there is no need to go beyond local equilibrium and assume additional relationships. This suggests, that the second law of thermodynamics should be reformulated, such that evolution of an irreversible system with non-zero thermodynamic inertia would be in agreement with it.

\eqr{eq/61} describes the entropy production, which is local not only in space, but also in time, i.e. \textit{instantaneous}. For the inertia-less system the instantaneous entropy production is a measure of instantaneous dissipation. However, for the system with thermodynamic inertia, the system response to a perturbation is not instantaneous. In particular, the instantaneous flux depends not on the instantaneous force, but on the force at time $\tau$ earlier. This leads to the observation, that the instantaneous entropy production is not very practical quantity to characterize dissipation in the system.

To illustrate this, let us consider the resistless system, described in \secr{sec/Resistless}. The evolution of the resistless system is a propagation of a heat wave. As every wave propagation, it is conservative. Still, this system obeys the property of local equilibrium, and the instantaneous entropy production for this system is given by \eqr{eq/61}. For the process of wave propagation, the instantaneous entropy production of the normal system oscillates around zero, taking both the positive and the negative values. Obviously, the instantaneous entropy production does not satisfy the second law of thermodynamics. Furthermore, according to the standard understanding of the entropy production, the wave propagation should be a process with alternating energy dissipation and energy gain. However, it is neither of these two: wave propagation is a conservative process. It follows therefore  that instantaneous entropy production has no practical meaning for the resistless system.

For the resistless system one can distinguish two kids of terms, which contribute to the Hamiltonian \eqref{eq/33}. They have the dimensionality of the entropy production, so we will refer to them as such. Namely, one can speak of the kinetic entropy production $m\,|\dot{\rho}|^2/2$ and $m\,|\dot{\rho}^{*}|^2/2$ and the potential entropy production $\ell\,|\nabla\varphi|^2/2$ and $\ell\,|\nabla\varphi^{*}|^2/2$. The kinetic entropy production exists due to the system's inertia $m neq 0$, while potential entropy production exists due to system's inhomogeneity $\nabla\varphi neq 0$. During the process of wave propagation the entropy is transferred from the kinetic term to the potential term and vise versa, keeping the Hamiltonians $\int\,\Hml_{\wave}\,dV$ and $\int\,\Hml_{\wave}^{*}\,dV$ constant. The instantaneous entropy productions \eqref{eq/61} also oscillate. This process is the same in both, the normal system and the mirror-image one.

In contrast, the dissipative inertia-less system does not have the kinetic term. Furthermore, the potential entropy production has the form of $\ell\,\nabla\varphi\cdot\nabla\varphi^{*}$, i.e. represents coupling between inhomogeneities in the normal and the mirror-image systems. In other words, the potential entropy production is shared between the normal and the mirror-image systems. Still, the Hamiltonian of the extended system $\int\,\widetilde{\Hml}_{\ic}\,dV$ is conserved. This makes the entropy to be transferred from the mirror-image system to the normal system. This process is irreversible, in contrast with the entropy transfer in the resistless system.

The extended system with non-zero inertia and non-zero resistance exhibit both processes of the entropy production transfer: reversible transfer between the kinetic and the potential term, as well as irreversible transfer between the mirror-image and the normal system, - such that the Hamiltonian of the extended system $\int\,\widetilde{\Hml}_{\ie}\,dV$ is conserved. It is clear, that the oscillatory part of the instantaneous entropy production is reversible and therefore does not contribute to dissipative behavior. 

In order to separate wave propagation from dissipation, we should consider the entropy production, which is averaged over the wave period $\tau_{\wave}$: 
\begin{equation}\label{eq/62}
\langle\sigma(\vR, t)\rangle_{\wave} \equiv \frac{1}{\tau_{\wave}}\int_{t}^{t+\tau_{\wave}}\,\sigma(\vR, t')\,dt'
\end{equation}
and $\langle\sigma^{*}(\vR, t)\rangle_{\wave}$ is introduced in the same way. Note, that the wave period $\tau_{\wave}$ is different from the delay time $\tau$. It depends on the system's size, as well as the speed of heat wave $c$. The statement of the second law of thermodynamics should then read, that the quantity which is positive in the course of evolution is the time-averaged entropy production of the normal system. Correspondingly, the time-averaged entropy production of the mirror-image system is negative:
\begin{equation}\label{eq/63}
\begin{array}{rl}
\langle\sigma(\vR, t)\rangle_{\wave} &\geq 0
\\\\
\langle\sigma^{*}(\vR, t)\rangle_{\wave} &\leq 0
\end{array}
\end{equation}

Substituting \eqr{eq/15} in \eqr{eq/62} we obtain the following expressions for the time-averaged local entropy productions:
\begin{equation}\label{eq/64}
\begin{array}{rl}
\langle\sigma\rangle_{\wave} =& r\,\left[\Delta_{\wave} (J^2) +\langle J^2\rangle \right]
\\\\
\langle\sigma^{*}\rangle_{\wave} =& r\,\left[\Delta_{\wave} ({J^{*}}^2) -\langle{J^{*}}^2\rangle\right]
\end{array}
\end{equation}
where $\Delta_{\wave} f(t) \equiv f(t+\tau_{\wave}) - f(t)$. It is convenient to examine the validity of \eqr{eq/63} for partial solutions. For the resistless system $r=0$ and both $\langle\sigma\rangle_{\wave} = 0$ and $\langle\sigma^{*}\rangle_{\wave} = 0$. For the steady heat wave caused by the oscillating external force or boundary conditions $J \propto \exp(i\omega t)$, and therefore $\Delta_{\wave} (J^2) = 0$. This leads to $\langle\sigma\rangle_{\wave} = r\,\langle J^2\rangle > 0$. Similarly, for the mirror-image system $\langle\sigma^{*}\rangle_{\wave} = -r\,\langle {J^{*}}^2\rangle < 0$. For the stationary process $J$ is independent of time, and therefore $\Delta_{\wave} (J^2) = 0$, which also leads to $\langle\sigma\rangle_{\wave} > 0$. Similarly, for the mirror-image system $\langle\sigma^{*}\rangle_{\wave} < 0$. Finally, for the relaxation process, which is the result of overdamping in the oscillatory system, $J \propto \exp(-\kappa t)$ and $\langle\sigma\rangle_{\wave} \propto 1-\kappa\tau$. In order for the time-averaged entropy production to be positive, we should have $\kappa\tau < 1$. However, this is exactly the condition of overdamping, which is true for the considered relaxation process by definition. Similarly, $J^{*} \propto \exp(\kappa t)$ and $\langle\sigma\rangle_{\wave} \propto -1+\kappa\tau < 0$. 

We note that one should not confuse the instantaneous entropy production, which is used in this paper in the context of mesoscopic continuous description, with the instantaneous entropy production which is formulated with the help of microscopic distribution function. Correspondingly, one should not confuse time averaging over the period of the heat wave with time averaging over the time of molecular equilibration. The analysis of this paper is performed withing the assumption of local equilibrium, which means that the system has already reached equilibrium locally, for a given position $\vR$ in space and a given moment $t$ in time. In particular, it is assumed that the time scale of molecular equilibration is much less than the period of the heat wave, so that molecular fluctuations have already been averaged at every moment $t$ in time.

\begin{figure}[ht!]
\centering 
\includegraphics[scale=0.5]{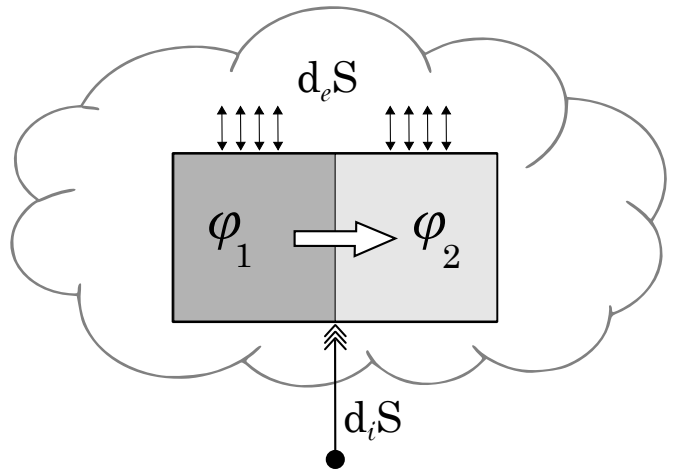}    
\caption{Conventional view on entropy transfers.}\label{fig-normal}
\end{figure}

Classical irreversible thermodynamics distinguishes between two types of entropy change \cite{deGrootMazur}. The first one, $d_eS = \delta Q/T$, is the equilibrium entropy change due to reversible interaction with the environment. The second one, $d_iS = \int\,\sigma\,dV$, is the production of entropy due to irreversible processes in the system. These transformations can be illustrated by \figr{fig-normal}, where we can see the entropy transformations in the systems, consisting of two thermal baths and the environment. The baths are in equilibrium with environment, but not in equilibrium with each other. Such system is open, i.e. not conservative, as the entropy is produced in the course of evolution.

\begin{figure}[ht!]
\centering 
\includegraphics[scale=0.5]{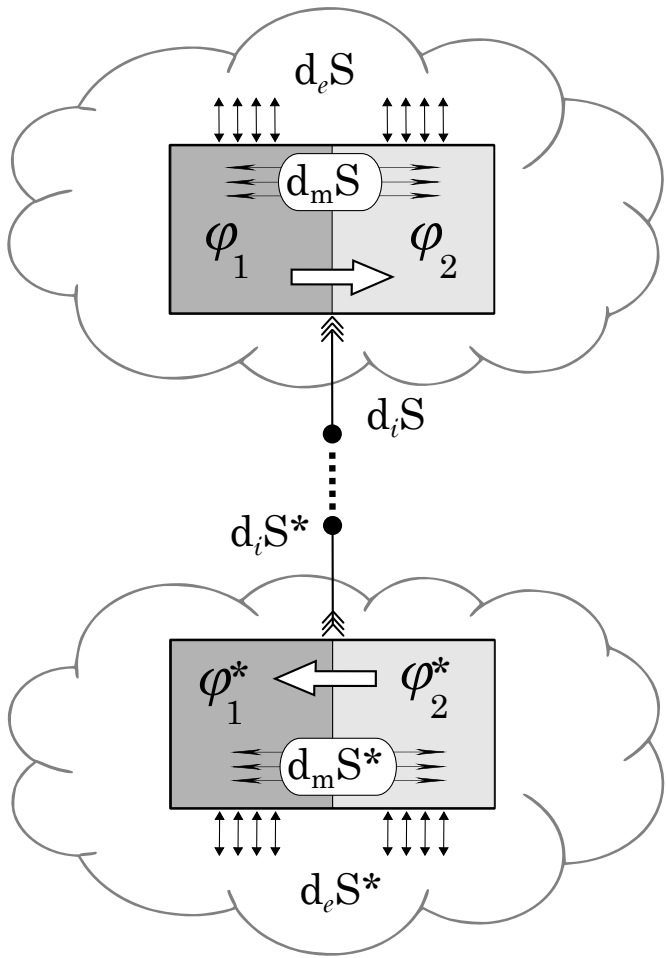}    
\caption{Entropy transfers in the extended system.}\label{fig-extended}
\end{figure}

The above analysis shows, that the entropy transformations in the irreversible system with non-zero thermodynamic inertia are different from those mentioned above. In particular, we can distinguish between three types of entropy change in the normal system and three types of the entropy change in the mirror-image system:
\begin{equation}\label{eq/65}
\begin{array}{rl}
dS =& d_eS + d_iS + d_mS
\\\\
dS^{*} =& d_eS^{*} + d_iS^{*} + d_mS^{*}
\end{array}
\end{equation}
The terms $d_eS$ and $d_iS$ in the entropy change of the normal system are the standard terms of classical irreversible thermodynamics. The third term, $d_mS$, is present due to non-zero thermodynamic inertia of the system. This term is reversible and accounts for the entropy transfer in the heat wave. Similarly, we can speak of the "classical" terms $d_eS^{*}$ and $d_iS^{*}$ in the mirror-image system. $d_eS^{*}$ is responsible for reversible exchange with environment, while $d_iS^{*}$ represent irreversible entropy destruction. In addition, there exists the third term $d_mS^{*}$, which has the similar nature as $d_mS$. In particular, it is reversible and responsible for the entropy transfer in the heat wave in the mirror-image system. These entropy transformations can be illustrated in \figr{fig-extended}. The extended system is closed. Indeed, the entropy transfer caused by the thermodynamic inertia is reversible and confined within each of the two subsystems, either the normal one or the mirror-image one. The irreversible entropy transfer happens \textit{from} the mirror-image system \textit{to} the normal system, i.e. is not produced or destroyed anywhere \textit{beyond} the extended system itself.
\section{Discussion}\label{sec/Discussion}

In this paper we considered evolution of the irreversible system with non-zero thermodynamic inertia. Behavior of such system has many analogies to the behavior of the systems, studied within well-established areas of physics, where inertia is important. In particular, these areas are classical mechanics and electrical engineering. 

\subsection{Classical mechanics}

The analogy with classical mechanics can be evident from the terminology used in the paper. Thus, for the resistless system the inertial entropy production $m\dot{\rho}^{2}/2$ is the direct analogy of the kinetic energy $m\dot{x}/2$ of a body with mass $m$ and position $x$, while the potential entropy production $\ell|\nabla\varphi|^2/2$ is the direct analogy of the potential energy $k(\Delta x)^2/2$ of the body at the position $\Delta x$ away from equilibrium with the force constant $k$. The potential energy of a body in classical mechanics is a function of its position only, and is independent of the velocity. Similarly, the potential entropy production is a function of the thermodynamic potential $\varphi$ only, and is independent of the its "speed" $\dot{\rho}$ or the flux $J$. Thus, the property of local equilibrium in irreversible thermodynamics is naturally satisfied in classical mechanics. The sum of the potential and the kinetic energy of an isolated mechanical system is conserved. Similarly, the sum of the potential and kinetic entropy production of a resistless thermodynamic system is conserved, which is illustrated, in particular, in \figr{fig-complex-resistless}. Furthermore, the analogy of the heat wave factor $\wave$ in classical mechanics is the eigenfrequency $\omega$. The wave factor does not have dimensionality of the inverse second, because we consider here a continuous system, which implies wave propagation, rather than oscillation of a single body. The typical characteristics of such system is the speed of wave propagation, $c$, which in our case is the speed of heat wave.

The thermodynamic system with non-zero resistivity is analogous to the mechanical system with non-zero friction. Conventional Hamiltonian formalism in mechanics is not applicable for such systems. However, it is still possible to account for friction by introducing the mirror-image system for a mechanical system as well \cite{MorseFeshbach}. The mirror-image system gains the energy dissipated from the normal system. Similarly, in irreversible thermodynamics, the mirror-image system gains the entropy, which is produced in the normal system, Indeed, the variational approach, which is applied in this paper, implies existence of a quantity, which has a minimum for the actual evolution, the action. For a thermodynamic system it has dimensionality of the entropy generation, and it is natural to understand the action as a kind of \textit{entropy generation} of the extended system (which consists of both, the normal one and the mirror-image one). The variational formalism implies the property, which can be formulated as following: \textit{the total entropy generation of the extended system during evolution reaches its minimum}. For the resistless system this leads to the conservation of the normal entropy, while for the dissipative system this leads to the entropy transfer from the mirror-image system to the normal system. The important consequence of this is that the normal and the mirror-image system should be considered together and some of their properties may be \textit{shared} within the extended system. The obvious example of the shared property is the Hamiltonian. Furthermore, we can speak of the potential entropy generation $\ell\nabla\varphi\nabla\varphi^{*}$ and the kinetic entropy generation $m\dot{\rho}\dot{\rho}^{*}$, sum of which is equal to the Hamiltonian. Thus, these terms may also be viewed as analogies to the potential and kinetic energy in classical mechanics.

The analogy between the mechanical and the thermodynamic quantities leads to analogy between the behavior of the mechanical and thermodynamic systems. A body with zero mass, which moves in media with friction, will eventually stop. If the body is attached to a spring and perturbed from its equilibrium position, it will stop in the spring equilibrium position, which corresponds to the minimum potential energy of the spring. This is a consequence of the so-called principle of minimum potential energy. Similarly, an irreversible thermodynamic system perturbed from equilibrium will experience relaxation. Eventually it will reach the equilibrium state, which corresponds to the state with maximum entropy. This process is formulated as the second law of thermodynamics. If follows therefore, that the second law of thermodynamics is the thermodynamic analogy of the mechanical principle of minimum potential energy. 

The principle of minimum potential energy is violated for systems with non-zero mass. Indeed, if the body has non-zero kinetic energy, its potential energy may increase. For the system without friction this leads to the decrease of the kinetic energy, such that the total energy remains the same. For the system with friction, the kinetic energy still can be transformed to the potential energy. The total energy, however, is not conserved, and the total energy averaged over the period of oscillations, decreases with time. Both systems, with and without friction, can be described by the variational formalism such that the energy is transferred between the kinetic and the potential form as well as between the normal and the mirror-image systems. Clearly, the principle of minimum potential energy is not directly applicable for the mechanical system with non-zero mass. Instead, one should use the principle of stationary action, either for a frictionless system or for a system with friction.

Similarly, the second law of thermodynamics is not directly applicable for a thermodynamic system with non-zero inertia. Instead, one should use the principle of stationary action, which has been formulated in this paper. This principle leads to conservation of the total entropy generation in the extended system.

\subsection{Electrical engineering}

The analogy with the electrical engineering is evident from the fact, that MVC equation has the form of the telegrapher's equation for an electrical transmission line. In particular, the pair of equation for the voltage $V$ and the current $I$ on the line, consisting the resistance $R$, the inductance $L$ and the capacitance $C$ is
\begin{equation}\label{eq/66}
\begin{array}{rl}
\nabla V =& -L\dot{I} - RI
\\\\
\nabla I =& -C \dot{V}
\end{array}
\end{equation}
The first of these equations is the analogy of MCV equation \eqref{eq/01} with $1/R$ being the thermal conductivity $\lambda$ and $L/R$ being the inertial parameter $\tau$, while the second of these equations is the analogy of the energy balance equation, with $C$ being the heat capacity. Furthermore, the analogy of the current $I$ is the flux $J$, while the analogy of the voltage $V$ is the temperature $T$. The current propagation in such systems has been extensively studied elsewhere. In particular, the inertial properties of the circuit are associated with a non-zero inductance $L$. Because of it the sinusoidal voltage and sinusoidal current have different phase: the AC voltage leads the AC current. If there is no external electromotive force, the current decays with time. The active resistance is the source of energy dissipation, while the presence of inductor and capacitor introduce the reactive resistance. It does not lead to dissipation of energy, but allow the elements to gain and release the energy in a reversible way.

The analysis above is directly applicable to such system. Since the temperature of the electrical circuit is considered to be constant and equal to the temperature of the environment, energy dissipation is proportional to the entropy dissipation. All the above relations and conclusions remain valid if we simply replace the word "entropy" with "energy". In particular, we should be able to distinguish between the kinetic and potential energy of the circuit. Furthermore, the impedance introduced in \secr{sec/Complex} is the electrical impedance, and the complex force-flux relations are the relations between the complex current and the complex voltage.

The power of the electric circuit is equal to the product of the current and the voltage. For the inertia-less circuit all power is positive and active, i.e. is released on the resistor. For the circuit with non-zero inertia the instantaneous power does not have much physical meaning. Because the current and the voltage are not in phase, the instantaneous power does not have definitive sign and oscillates. One can distinguish between the active power, which is released on the resistor, and the reactive power, which characterizes the oscillations of the electric current. To discard the power oscillations during one period of AC, it is useful to consider the power, which is averaged over the period of AC. The averaged power indicates the losses on the resistor only, and therefore is always positive.

Comparing the behavior of the electric circuit and the inertial irreversible system, we can observe that the electric power is the analogy of the entropy production. For DC current or for an inertia-less circuit the instantaneous electric power is always positive, which is the statement of the second law of thermodynamics. In contrast, for AC current in the circuit with inertia it is only the time-averaged power which is positive, while the instantaneous power does not have definitive sign. The same applies to the entropy production in the electric circuit and in the irreversible system with inertia in general.

\subsection{Concluding remarks}

Let us summarize the findings presented in this paper. We have shown that irreversible evolution of a thermodynamic system can be derived from a variational principle of stationary action. In order to formulate this principle we need to consider the extended system, which consists of the normal system and the mirror-image system. The action, which is being minimized has the dimensionality of entropy and can be viewed as the entropy generation of the extended system. Thus, the principle of stationary action states that evolution of the extended system minimizes the generated entropy of the extended system.

The Lagrangian introduced in this paper describes general evolution of an irreversible system. It can also describe the specific cases: resistless evolution of the system with non-zero inertia, which leads to wave propagation; as well as dissipative evolution of inertia-less system, which leads to relaxation. The evolution equations or the force-flux relations are derived from the variational procedure. They represent the Euler-Lagrange equations for the corresponding Lagrangian. The important part of the analysis is the property of local equilibrium, which states that the non-equilibrium relation between the material density and the potential has the same form as in equilibrium. It appears, that the property of local equilibrium is sufficient for derivation of the experimentally verified evolution equations. Thus, no additional relations which characterize non-equilibrium evolution are needed.

Earlier it has been shown that dissipative evolution of inertia-less system results in the second law of thermodynamics. It states that instantaneous entropy production of the non-equilibrium normal system is always positive. Here we have shown that this formulation is no longer valid for the dissipative system with non-zero inertia. However, one can modify the statement of the second law of thermodynamics. In particular, it should read: for the irreversible evolution of the system the averaged over the period of the heat wave entropy production is always positive.

%
\bibliographystyle{unsrt}

\end{document}